\documentclass[preprint,showpacs,preprintnumbers,amsmath,amssymb,nofootinbib]{revtex4}

\usepackage{graphicx}
\usepackage{dcolumn}
\usepackage{bm}

\newcommand{\dis}[1]{\begin{equation}\begin{split}#1\end{split}\end{equation}}
\newcommand{\beqa}[1]{\begin{eqnarray}#1\end{eqnarray}}

\newcommand{\bi}{\bibitem}
\newcommand{\nn}{\nonumber}

\newcommand{\be}{\begin{eqnarray}}
\newcommand{\ee}{\end{eqnarray}}
\catcode`\@=11
\def\lsim{\mathrel{\mathpalette\@versim<}}
\def\gsim{\mathrel{\mathpalette\@versim>}}
\def\@versim#1#2{\vcenter{\offinterlineskip
\ialign{$\m@th#1\hfil##\hfil$\crcr#2\crcr\sim\crcr } }}
\catcode`\@=12

\begin{document}

\preprint{MPP-2004-15}

\title{Double Suppression of 
FCNCs in Supersymmetric Models}

\author{Ki-Young Choi$^{1,2}$}
\author{Yuji Kajiyama$^{3}$}
\author{Jisuke Kubo$^{3,4}$}
\author{Hyun Min Lee$^{2}$}

\affiliation{
$^{1}$School of Physics and Center for Theoretical Physics,
Seoul National University,
Seoul 151-747, Korea\\
$^{2}$Physikalisches Institut,
Universit\" at Bonn,
D-53115 Bonn, Germany\\
$^{3}$Institute for Theoretical Physics, 
Kanazawa University, Kanazawa 920-1192, Japan\\
$^{4}$ Max-Planck-Institut f\"ur Physik,
 Werner-Heisenberg-Institut,
D-80805 Munich, Germany
 }

\vspace{2cm}
\begin{abstract}
A mechanism for double suppression of 
flavor-changing neutral currents (FCNCs) and CP violating phases in
supersymmetric  models
is suggested.
At $M_{\rm SUSY}$ they are suppressed due to
a nonabelian  discrete flavor symmetry,
and the infrared attractive force of gauge interactions
in extra dimensions are  used to
suppress them at the compactification scale.
We present a concrete model, which is 
a simple extension of $S_{3}$ invariant
minimal supersymmetric standard model, 
where only $SU(2)_{L}$ and $SU(3)_{C}$ gauge multiplets
are assumed to propagate in the bulk.
We find that a disorder of two orders of magnitude
in soft supersymmetry breaking parameters
above the compactification
scale may be allowed
to satisfy experimental constraints on FCNC processes
and CP violating phenomena.

\end{abstract}


\maketitle

\section{Introduction}

Low energy supersymmetry  (SUSY) is introduced to protect
the Higgs mass from the quadratic divergence \cite{susy}.
Since low energy SUSY is  broken, the breaking of SUSY
must be soft, whatever its origin is, 
to maintain the very nature of low energy SUSY.
Unfortunately, 
the most arbitrary part of a phenomenologically
viable supersymmetric extension
of the standard model (SM) is this soft breaking sector, 
because renormalizability allows an introduction of
many  soft supersymmetry breaking (SSB) parameters.
In the minimal supersymmetric standard model (MSSM), 
 more than 100 SSB  parameters can be introduced \cite{dimopoulos1}.
The problem is not only this large number of
the SSB parameters, but also the fact that one has to
highly fine tune them so that they do not 
induce unacceptably 
 large flavor changing neutral currents (FCNCs) and
CP violations \cite{fcnc-mueg,fcnc-k,fcnc-edm,fcnc-bsg,fcnc}.
This problem,
called the SUSY flavor problem, is not new, but has existed ever
since supersymmetry found phenomenological 
applications \cite{dimopoulos2}. 

There are several theoretical approaches
to overcome this problem, which may be 
divided in three types.
In the case of the first type it is assumed
 that there exists a hidden sector in which
 SUSY is broken by some flavor blind mechanism,
 and that  SUSY breaking is mediated by  flavor blind interactions
 to the MSSM sector \cite{gauge,anomaly,gaugino,susy}.
The idea of the second type to overcome the SUSY flavor problem is to
use the infrared attractive force of
 gauge interactions
\cite{ross, karch,ns,knt,ls,abel,kubo6,choi1,choi2,kaji1}.
It has been found in \cite{kubo6,choi1,choi2,kaji1}
that thanks to the power-running law of
gauge couplings \cite{veneziano,dienes1,ejiri}
 in extra dimensions \cite{antoniadis1,arkani1},
the infrared attractiveness of running 
SSB parameters \cite{kobayashi1} 
 can be so amplified that at a compactification scale
the SSB terms align themselves out of their anarchical disorder at a
cutoff  scale $\Lambda$, even if the ratio of the cutoff scale
$\Lambda$ to the compactification scale $\Lambda_{C}$ is small
 $\sim O(10^3)$.
The third mechanism is based on a flavor symmetry
principle \cite{hall2,hamaguchi,babu1,kobayashi3,king2}.
In this approach one should be aware of the fact that,  if
a flavor symmetry is hardly 
broken  at low energies e.g. in the Yukawa sector,
then  these interactions  can induce
non-symmetric SSB terms.
In \cite{king2} it has been shown that a spontaneously
broken continuous  horizontal symmetry  based 
on $SU(3)$ \cite{king1} can significantly suppress
FCNCs and CP violating phases.
An advantage of a discrete flavor symmetry is that
no Nambu-Goldstone bosons can occur when it is spontaneously
broken. In \cite{hamaguchi,kobayashi3} it has been found that
a discrete flavor symmetry based on $S_{3}$ at 
low energy
can considerably soften the SUSY flavor problem.

In this paper, we are motivated by the
desire to combine  the second and third types of mechanism  to 
soften  the SUSY flavor problem.
That is, in this double suppression
mechanism,
asymptotically free gauge interactions
in extra dimensions  bring
a large  disorder  of the SSB parameters 
at the cutoff scale $\Lambda$ down
to $O(1)$ disorder at the compactification scale $\Lambda_{C}$, and
FCNCs and CP phases, which are induced by
the SSB parameters in $O(1)$ disorder at  low energy,
are  suppressed due to  an intact flavor  symmetry.
By $O(1)$ disorder of the SSB parameters we mean e.g.
$(m^{2}_{\rm max}-m^{2}_{\rm min})/m^{2}_{\rm av} \sim O(1)$,
where $m^{2}_{\rm max, min,av}$ are 
the maximal, minimal  and average squared soft scalar masses, respectively, and 
similarly for the other SSB parameters.

In Sect. II we  start by  considering a minimal supersymmetric extension
of the $S_{3}$ invariant SM of \cite{kubo1,kubo2}. 
In \cite{kobayashi3}, such an extension has been indeed made.
There, additional $SU(2)_{L}\times U(1)_{Y}$ singlet Higgs
multiplets have been introduced to avoid the appearance
of pseudo Nambu-Goldstone bosons.
Here we would like to propose an alternative way
to avoid the appearance
of the pseudo Nambu-Goldstone bosons.
We assume that $S_{2}$ is softly broken by
certain $B$ terms in the SSB sector.
It  turns out that it is possible  to make all the Higgses except for one 
very heavy  without running into the 
problem of triviality. 

In Sect. III we  discuss  the third type of suppression
mechanism.
We explicitly compute
$\delta^{q,\ell}_{LL,RR,LR}$ of \cite{fcnc} in terms of the $S_{3}$
invariant SSB terms, and find that
these $\delta$'s can satisfy the experimental constraints
if the disorder of the  SSB parameters at $M_{\rm SUSY}$
is at most about one.
The analysis of this section is basically a reanalysis of \cite{kobayashi3}.
We, however,  use a new set of input parameters in the quark sector
to obtain a better agreement  with the 
experimental results on the Cabibbo-Kobayashi-Maskawa mixing
matrix $V_{\rm CKM}$.

Extra dimensions are introduced in Sect. IV.
Our model  is a simple extension 
 into $\delta$ extra dimensions along 
 the line of \cite{antoniadis1,arkani1,dienes1}, in which 
it is assumed that 
the matter multiplets and the $U(1)_Y$ gauge multiplet are located 
at a fixed point, while the $SU(2)_L$ and $SU(3)_C$ gauge multiplets propagate
in the bulk. 

Sect. V is devoted to conclusion, and in Appendix
we give  renormalization group (RG) functions that are used in this paper.

\section{$S_{3}$ invariant extension of the MSSM}

Flavor symmetries based on a 
permutation symmetry have been considered 
by many authors in the past \footnote{One of the first papers
on permutation symmetries  are
\cite{pakvasa1,harari,koide1}.
See \cite{kubo2} for a partial list of references, and
\cite{fritzsch} for a review.
By a flavor symmetry we mean
 a symmetry of the Yukawa interactions, because
 a symmetry of a mass matrix is not necessarily
a symmetry of the theory. }.
Phenomenologically
viable models based on 
non-abelian discrete flavor symmetries $S_{3}, D_{4}$ and $A_{4}$ 
and also on a product of abelian discrete symmetries 
 have been recently constructed in
 \cite{kubo1,kubo2},  \cite{grimus2},  
 \cite{ma1,babu2,hirsch} and \cite{grimus1,ma4}, respectively,
which can  naturally explain a large neutrino mixing.
(See also  \cite{koide2,ohlsson,kitabayashi,ma4,grimus3}. )
In this section  we would like to consider a supersymmetric extension 
of the $S_{3}$ invariant model of \cite{kubo1,kubo2}.

\subsection{$S_{3}$ invariant superpotential }
Three generations of the quarks and leptons
 belong to the reducible representation
of $S_3$ ${\bf 3}={\bf 1}+{\bf 2}$, respectively.
They are denoted by
$Q_{I},Q_{3},U_{IR},U_{3R},D_{IR},D_{3R},
L_{I},L_{3},E_{IR},\\E_{3R},N_{IR},N_{3R}$
in an obvious notation.
We also introduce an $S_3$ doublet Higgs pair,
$H^U_I, H^D_I (I=1, 2)$, as well as an
$S_3$ singlet Higgs pair, $H^U_3, H^D_3$.
The same R-parity is assigned to these fields as in the MSSM.
Then we assume that the superpotential,
$W = W_D + W_U + W_E + W_{N}+W_{M}+W_{H}$, is
invariant under  $S_3$ permutations.
Each part is given explicitly as follows:
\be
W_D &=&
Y_1^D Q_I H^{D}_3 D_{IR} + Y_3^D Q_3 H^{D}_3 D_{3R}  \nn \\
& &+ Y^{D}_{2}f_{IJK} Q_{I}  H^{D}_{J}  D_{KR}
\nn\\
& & + Y^D_{4} Q_3 H^D_I  D_{IR}
+ Y^D_{5} Q_I H^D_I D_{3R},
\label{potd}\\
W_U &=&
Y_1^U Q_I H^{U}_3 U_{IR} + Y_3^U Q_3 H^{U}_3 U_{3R}  \nn \\
& &+ Y^{U}_{2}f_{IJK} Q_{I}  H^{U}_J  U_{KR} \nn\\
& & + Y^U_{4} Q_3 H^U_I  U_{IR}
+ Y^U_{5} Q_I H^U_I U_{3R},
\label{potu}
\ee
\be
W_E &=&
Y_1^E L_I H^{D}_3 E_{IR} + Y_3^E L_3 H^{D}_3 E_{3R}  \nn\\
& &+Y^{E}_{2}f_{IJK}L_{I} H^{D}_J  E_{KR}\nn\\
& & + Y^E_{4} L_3 H^D_I  E_{IR}
+ Y^E_{5} L_I H^D_I E_{3R},
\label{pote}\\
W_N &=&
Y_1^N L_I H^{U}_3 N_{IR} + Y_3^N L_3 H^{U}_3 N_{3R}  \nn\\
& &+Y^{N}_{2}f_{IJK}L_{I} H^{U}_J  N_{KR}\nn\\
& & + Y^N_{4} L_3 H^U_I  N_{IR}
+ Y^N_{5} L_I H^U_I N_{3R},
\label{potn}\\
W_{M} &=& \frac{1}{2}M_{1}N_{IR}N_{IR}+ \frac{1}{2}M_{3}N_{3R}N_{3R},\\
\label{maj-mass}
W_H &=&
\mu_{1} H^{U}_{I}H^{D}_{I}+\mu_{3} H^{U}_{3}H^{D}_{3},
\label{poth}
\ee
where
\be
f_{121} &=& f_{211}=f_{112}=-f_{222}=1~,~
f_{111}=f_{221}=f_{122}=f_{212}=0.
\label{fijk}
\ee
($f_{IJK}$ is completely symmetric).
$W$ is the most general renormalizable superpotential
that is $S_{3}\times R$ invariant.

\subsection{Soft supersymmetry breaking sector}
\noindent
(i) Gaugino masses:\\
The gaugino masses are the same as in the MSSM.

\vspace{0.3cm}
\noindent
(ii) Trilinear couplings:\\
The trilinear couplings $h$'s can be read off from
$W_{U,D,E,N}$.
We denote them
by $h_{I}^{U}$ etc.
By symmetry, the trilinear couplings 
have exactly the same structure as the Yukawa couplings.

\vspace{0.3cm}
\noindent
(ii)  Soft scalar masses:\\
$S_{3}$ invariant soft scalar masses are {\em diagonal}, and
have the general structure:
\be
\tilde{m}_{1}^{2}
\hat{\phi}_{I}^{*}\hat{\phi}_{I}+
\tilde{m}_{3}^{2}\hat{\phi}_{3}^{*}\hat{\phi}_{3}
\label{softm2}
\ee
for all scalar components $\hat{\phi}$.
Those of the MSSM are denoted by
$m_{Q_{I}}^{2},m_{Q_{3}}^{2},m_{U_{IR}}^{2},m_{U_{3R}}^{2}$
etc.

\vspace{0.3cm}
\noindent
(iii)  B-terms:\\
$S_{3}$ invariant B-terms are
\be
{\cal L}_{B} &=&B_{1}(\hat{H}_{1}^{U}\hat{H}_{1}^{D}+
\hat{H}_{2}^{U}\hat{H}_{2}^{D})+B_{3}(\hat{H}_{3}^{U}\hat{H}_{3}^{D})
+h.c.
\label{bterm1}
\ee

Given the superpotential (\ref{potd})-(\ref{poth}) along with the SSB sector,
we can now write down the scalar potential.
For simplicity we assume that only the neutral scalar components
of the Higgs supermultiplets acquire VEVs.
The relevant part of the scalar potential is then given by
\be
V &=&
(|\mu_{1}|^{2}+m_{H_{1}^{U}}^{2})(|\hat{H}_{1}^{0U}|^{2}+
|\hat{H}_{2}^{0U}|^{2})+
(|\mu_{1}|^{2}+m_{H_{1}^{D}}^{2})(|\hat{H}_{1}^{0D}|^{2}
+|\hat{H}_{2}^{0D}|^{2})\nn\\
& &+(|\mu_{3}|^{2}+m_{H_{3}^{U}}^{2})(|\hat{H}_{3}^{0U}|^{2})+
(|\mu_{3}|^{2}+m_{H_{3}^{D}}^{2})(|\hat{H}_{3}^{0D}|^{2})\nn\\
& &+\frac{1}{8}(\frac{3}{5}g_{1}^{2}+
g_{2}^{2})(|\hat{H}_{1}^{0U}|^{2}+
|\hat{H}_{2}^{0U}|^{2} +|\hat{H}_{3}^{0U}|^{2}
-|\hat{H}_{1}^{0D}|^{2}-
|\hat{H}_{2}^{0D}|^{2} -|\hat{H}_{3}^{0D}|^{2} )^{2}\nn\\
& &-[~B_{1}(\hat{H}_{1}^{0U}\hat{H}_{1}^{0D}+
\hat{H}_{2}^{0U}\hat{H}_{2}^{0D})+
B_{3}(\hat{H}_{3}^{0U}\hat{H}_{3}^{0D})
+h.c.~]
\label{scalarp1}
\ee
As one can see easily, the scalar potential $V$ (\ref{scalarp1})
has a continuous global symmetry $SU(2)\times U(1)$ in addition
to the local electroweak gauge symmetry 
$SU(2)_{L}\times U(1)_{Y}$.
As a result, there will be a number of
pseudo Goldstone bosons that are phenomenologically
unacceptable. This is a consequence of $S_{3}$ symmetry.
(A similar consequence exists also in nonsupersymmetric case, too.)
Therefore, we have to break $S_{3}$ symmetry explicitly.
We would like to break it
as soft as possible to 
preserve predictions from $S_{3}$ symmetry, while
breaking the global  $SU(2)\times U(1)$
symmetry completely.
Surprisingly, there is a unique choice for a set of soft $S_{3}$ breaking 
terms: The softest terms
in the present case have the canonical  dimension  two,
implying they should be in the SSB sector.
As for the soft scalar masses, we have an 
important consequence (\ref{softm2}) from $S_{3}$ symmetry that
they are diagonal in generations.
Since we would like to preserve this structure, the only choice is
to introduce  the soft $S_{3}$ breaking 
terms in  the B sector (\ref{bterm1}).
Moreover, looking at the $S_{3}$  invariant scalar potential $V$ (\ref{scalarp1}),
we observe that it has an abelian discrete symmetry
\be
S_{2}' &:& \hat{H}_{1}^{U,D} \leftrightarrow \hat{H}_{2}^{U,D}.
\label{s2p}
\ee
We assume that the soft $S_{3}$ breaking 
terms respect
this discrete symmetry (\ref{s2p}), and add the following
 soft $S_{3}$ breaking Lagrangian:
\be
{\cal L}_{S_{3}B} &=&
B_{4}(\hat{H}_{1}^{U}\hat{H}_{2}^{D}
+\hat{H}_{2}^{U}\hat{H}_{1}^{D})
+B_{5}\hat{H}_{3}^{U}(\hat{H}_{1}^{D}+\hat{H}_{2}^{D})
+B_{6}\hat{H}_{3}^{D}(\hat{H}_{1}^{U}+\hat{H}_{2}^{U})+h.c.
\label{s3b}
\ee
The resulting scalar potential can be analyzed, and one finds that
a local minimum respecting 
$S_{2}'$  symmetry, i.e.
\be 
<\hat{H}_{1}^{U}> &=&<\hat{H}_{2}^{U}>=v_{U}/2\neq 0~,~
<\hat{H}_{1}^{D}> =<\hat{H}_{2}^{D}>=v_{D}/2\neq 0,\nn\\
<\hat{H}_{3}^{U}> &=& v_{3U}/\sqrt{2}\neq 0~,~
<\hat{H}_{3}^{D}>= v_{3D}/\sqrt{2}\neq 0,
\label{vevs}
\ee
can occur.
We find the
the lightest Higgs, the SM Higgs, can be written as a linear combination
\be
h_{\rm SM} &=&
(v_{D}\hat{H}_{+}^{0D}+v_{3D}\hat{H}_{3}^{0D}+
v_{U}\hat{H}_{+}^{0U}+v_{3U}\hat{H}_{3}^{0U})/v,
\ee
where
\be
\hat{H}_{+}^{0D,U} &=&\frac{1}{\sqrt{2}}(
\hat{H}_{1}^{0D,U}+\hat{H}_{2}^{0D,U})~,~
v=(v_{U}^{2}+v_{3U}^{2}+v_{D}^{2}+v_{3D}^{2})^{1/2}
\simeq 246 ~\mbox{GeV}.
\ee
Its mass is approximately given by
\be
m_{h} &\simeq &((3/5)g_{1}^{2}+g_{2}^{2})
(v_{U}^{2}+v_{3U}^{2}-v_{D}^{2}-v_{3D}^{2})^{2}/v
\label{mh}
\ee
for $\mu^{2\prime}s , B's >> v^{2}$. It can be shown that the masses 
of  other Higgs multiplets can be made
arbitrarily heavy. 
From (\ref{mh}) we see that the tree-level upper bound
for $m_{h}$ is exactly the same as in the MSSM.

Because of the very nature of the SSB terms, the explicit breaking of $S_{3}$
in the B terms (\ref{s3b})
does not propagate to the other sector
in the sense that it does not produce 
 $S_{3}$ violating  infinities in  other sectors.
Furthermore, although the superpotential (\ref{potd})-(\ref{poth})
and the corresponding trilinear couplings 
do not respect the $S_{2}'$ symmetry (\ref{s2p}),
they can not generate $S_{2}'$
violating infinite B terms  because 
they  can generate only $S_{3}$ invariant
terms in the Higgs sector, 
which are however automatically $S_{2}'$ invariant.

\subsection{$Z_{2}$ symmetry in the leptonic sector,
mass matrices and diagonalization}
As in \cite{kubo1,kubo2}
 we assume the existence of an abelian discrete symmetry $Z_{2}$
in the leptonic sector.
The $Z_2$ parity assignment is:
\be
+&  & \mbox{ for} ~~H_I^{U,D}, ~L_3, ~L_I, ~E_{3R},~ E_{IR},~
N_{IR}
~\mbox{ and}~~
- ~~\mbox{for}~~H_3^{U,D},~N_{3R},
\label{z2}
\ee
which kills
\be
Y_{1}^{E},Y_{3}^{E},Y_{1}^{N},~\mbox{and}~Y_{5}^{N}
\ee
in the Yukawa sector, and
\be
h_{1}^{E},h_{3}^{E},h_{1}^{N},~\mbox{and}~h_{5}^{N}
\ee
in the SSB sector.
Note that this symmetry is explicitly broken in the quark sector.
However, the $F$ term non-renormalization theorem
prevents from producing $Z_{2}$ violating terms in the leptonic sector.
As a result of (\ref{vevs}), the quark and lepton mass matrices
take the general form
\be
{\bf M}_a = \left( \begin{array}{ccc}
m^a_1+m^a_{2} & m^a_{2} & m^a_{5}
\\  m^a_{2} & m^a_1-m^a_{2} &m^a_{5}
    \\ m^a_{4} & m^a_{4}&  m^a_3
\end{array}\right) ~
\mbox{with}~m_{1}^{e}=m_{3}^{e}=m_{1}^{\nu}=m_{5}^{\nu}=0,
\label{general-m}
\ee
where $a=u, d, e,\nu $. It has been found in \cite{kubo1}
 that the mass matrices of the  general form
(\ref{general-m}) are  consistent with all the observed
quark and lepton masses and mixing angles.
We emphasize that this result can remain valid after supersymmetrization of the
model.

Next we consider diagonalization of the mass matrices, and start
with the charged lepton mass matrix ${\bf M}_e$.
It has only   real parameters, and they are given by
\be
m_{2}^{e} &\simeq & 0.07474~\mbox{GeV}, 
m_{4}^{e} \simeq 0.0005141~\mbox{GeV}, 
m_{5}^{e} \simeq  1.254~\mbox{GeV}, m_{1}^{e}=m_{3}^{e}=0.
\ee
Then the unitary matrices 
$U_{eL}$ and $U_{eR}$
defined as
\be
U_{eL}^{\dagger} {\bf M}_e U_{eR}
= \mbox{diag}( m_e, m_{\mu}, m_{\tau})
\ee
are found to be
\be
U_{eL} & \simeq &\left(
\begin{array}{ccc}
-0.003452 & -0.7071 & 0.7071\\
0.003427 & 0.7071 & 0.7071\\
1.0 & -0.004864 & 0.00001722
\end{array}\right),
\label{UeL}
\\
U_{eR} & \simeq  &\left(
\begin{array}{ccc}
0.9982 & -0.000023578 & 0.05949\\
0.00002362 & -1.0  & 0\\
-0.05949 & 1.4 \times 10^{-6} &0.9982
\end{array}\right).
\label{UeR}
\ee
Note that the maximal mixing appearing in $U_{eL}$
 is responsible for the atmospheric neutrino
mixing, while
the solar neutrino mixing is explained by the
large mixing in the rotation matrix of the neutrino mass matrix \footnote{
See \cite{smirnov} for recent developments on neutrino physics.}.
That is, the elements,  $(U_{eL})_{21}$ and $(U_{eL})_{23}$
become, respectively, the $(1,3)$
and $(3,3)$ elements of the neutrino mixing matrix $V_{\rm MNS}$;
\be
U_{e3} & \simeq &-0.0034~,~
\cos\theta_{\rm atm}=1/\sqrt{2}.
\ee

As in  the leptonic case, we introduce  unitary
matrices $U_{u(d)L}$ and  $U_{u(d)R}$ satisfying
\be
U_{u(d)L}^{\dagger} {\bf M}_{u(d)} U_{u(d)R} =
\mbox{diag} (m_{u(d)}, m_{c(s)}, m_{t(b)}).
\ee
Realistic quark masses 
as well as 
CKM matrix $V_{\rm CKM}$ can be obtained from \footnote{The values
of the input parameters used here are slightly different from
those in \cite{kubo1,kobayashi3}.
They yield 
$V_{\rm CKM}$ in a better agreement with the experimental values.}
\be
m_1^u &=& - 0.06504~\mbox{GeV}~,~ m_2^u = - 0.06148~\mbox{GeV}~,~
m_3^u =173.5 ~\mbox{GeV},\nn\\
m_4^u &=&10.70 ~\mbox{GeV} ~,~
m_5^u=4.166 ~\mbox{GeV}\nn \\
m_1^d &=&0.008974~\mbox{GeV}~,~  
m_2^d =0.01460~\mbox{GeV}~,~
m_3^d =1.950 + 1.548 I ~\mbox{GeV},\nn\\
m_4^d& =& 1.045 ~\mbox{GeV}~,~
m_5^d =0.1427  ~\mbox{GeV},
\label{choice1}
\ee
where we have assumed that only $m_3^d$
is complex. The values given in (\ref{choice1}) are
not a unique choice. However, numerical analyses
show that the values can not be continuously deformed without
changing the values of the quark masses and $V_{\rm CKM}$
(under the assumption that only $m_{3}^{d}$ is complex).
The  unitary matrices for the set of the parameter values (\ref{choice1}) are given by
\be
 U_{uL}   &\simeq & \left(
\begin{array}{ccc}
0.6366 & -0.7709 & 0.02375 \\
 -0.7712 &-0.6361 & 0.02379\\
 0.003233 & 0.03346 & 0.9994
\end{array}\right),
\label{UuL}
\\
U_{uR} & \simeq   &\left(
\begin{array}{ccc}
 0.6365 & 0.7688 & 0.06136\\
-0.7712 & 0.6336 & 0.06138\\
0.008306 & -0.08639 & 0.9962
\end{array}\right),
\label{UuR}
\ee
\be
U_{dL}  &\simeq& \left(
\begin{array}{ccc}
 -0.7497,0.549,  0.04444\\
0.6284,0.7409, 0.0410\\
0.01124,-0.06039, 0.9424
\end{array}\right)\nn\\
 & & + I \left(
\begin{array}{ccc}
 0.2037,- 0.3043, - 0.01238\\
 - 0.03858,- 0.2299, - 0.01357\\
 - 0.00321, - 0.01458,0.3285
\end{array}\right), 
\label{UdL} \\
U_{dR}  &\simeq& \left(
\begin{array}{ccc}
0.74582,-0.4371, 0.3400\\
-0.6277,-0.6584, 0.3396\\
-0.08189,0.4862, 0.8129
\end{array}\right)\nn\\
 & & + I \left(
\begin{array}{ccc}
0.2027,- 0.2863,0.1182\\
 - 0.03772,- 0.2043,0.1183\\
 - 0.02341, - 0.1238, - 0.2834
\end{array}\right). 
\label{UdR}
\ee
Note that the off-diagonal elements in $U_{dL}$ and $U_{dR}$ carry
large complex phases.
For completeness we write 
the explicit values of $V_{\rm CKM}$ and the quark masses 
for the set of parameters (\ref{choice1}):
\be
|V_{\rm CKM} | &=|U_{uL}^{\dag}U_{dL}|& \simeq \left(
\begin{array}{ccc}
0.975,0.223,0.0037\\
0.222,0.974,0.041\\
 0.0084,0.040,0.999
\end{array}\right), 
\label{ckm}
\ee
with
\be
J &\simeq & 3.0\times 10^{-3}~,~
\phi_{1}(\beta) \simeq   23^o~,~ \phi_{3}(\gamma) \simeq  67^{o},
\ee
and 
\be
m_{d}  & \simeq & 4.4 ~\mbox{MeV}~,~
m_{s} \simeq 0.09 ~\mbox{GeV}~,~
m_{b}  \simeq  2.9 ~\mbox{GeV},\nn\\
m_{u} & \simeq & 2.3 ~\mbox{MeV}~,~
m_{c} \simeq 0.64~\mbox{GeV}~,~
m_{t} \simeq  174~\mbox{GeV},
\ee
where $J$ is the Jarlskog invariant.

\section{Suppression of FCNCs at $M_{\rm SUSY}$}
Since all the soft scalar masses have the form 
(\ref{softm2}), we 
write the mass matrices as
\be
{\bf \tilde{m}^2}_{aLL} =
m^2_{a} \left(
\begin{array}{ccc}
a_L^{a} & 0 & 0 \\
0 & a_L^{a} & 0 \\
0 & 0 & b_L^{a}
\end{array}
\right),~
{\bf \tilde{m}^2}_{aRR} =
m^2_{a} \left(
\begin{array}{ccc}
a_R^{a} & 0 & 0 \\
0 & a_R^{a} & 0 \\
0 & 0 & b_R^{a}
\end{array}
\right)~~~(a=\tilde{\ell},\tilde{q}),
\label{scalarmass}
\ee
where $m_{\tilde{\ell},\tilde{q}}$ denote the average of the
slepton and squark masses, respectively,  and $(a_{L(R)}, b_{L(R)})$ are
dimensionless free parameters of $O(1)$.
Further, since the trilinear  interactions 
are also  $S_{3}$ invariant,
the left-right mass matrix can  be written as
\be
{\bf \tilde{m}^2}_{aLR} =  \left(
\begin{array}{ccc}
m_1^{a} A_1^{a}+m_2^{a} A_2^{a}
 & m_2^{a} A_2^{a} & m_5^{a} A_5^{a} \\
m_2^{a} A_2^{a} & m_1^{a} A_1^{a}-m_2^{a} A_2^{a}
 & m_5^{a} A_5^{a} \\
m_4^{a} A_4^{a} & m_4^{a} A_4^{a} & m_3^{a} A_3^{a}
\end{array}
\right)~~~(a=\tilde{\ell},\tilde{q}),
\ee
where $A_i^{a}$ are free parameters of dimension one. Here we 
assume that they are in  the same order as the gaugino masses.

We consider FCNC processes, {\it e.g.}
$Br(\mu \rightarrow e + \gamma)$, that 
are proportional to the off-diagonal elements of
\be
\Delta_{LL,RR}^{a} &=&
U_{aL,R}^{\dagger} ~{\bf \tilde{m}^2}_{aLL,RR}~ U_{aL,R}~\mbox{and}~
\Delta_{LR}^{a} =
U_{aL}^{\dagger}~ {\bf \tilde{m}^2}_{aLR} ~U_{aR}.
\ee
By using the unitary matrices  given in
Eqs.~(\ref{UeL}),(\ref{UeR}) and (\ref{UuL})-(\ref{UdR}),
$\Delta$'s  can be explicitly evaluated.
In \cite{fcnc}, experimental bounds  on the dimensionless quantities
\be
\delta^{a}_{LL,RR,LR} &=& \Delta^{a}_{LL,RR,LR}/m^2_{\tilde{a}}~~~(a=\ell,q),
\ee
are given, which are summarized in Table I.
The theoretical values of $\delta$'s for the present model 
are calculated below,
where 
\be
\Delta a_{L,R}^{a} &=&a_{L,R}^{a}-b_{L,R}^{a},~
\tilde{A}_{i}^{a} = \frac{A_{i}^{a}}{m_{\tilde{a}}}~~(a=\ell,q),
\label{deltaAt1}
 \ee
 and $a_{L,R},b_{L,R}$ are defined in (\ref{scalarmass}).

\vskip 0.5cm
\underline{\bf Leptonic sector ($LL$ and $RR$): }
\be
(\delta^{\ell}_{12})_{LL}
&\simeq& (\delta^{\ell}_{21})_{LL}\simeq
 4.8 \times 10^{-3} ~\Delta a_L^{\ell}, \nn\\
(\delta^{\ell}_{13})_{LL}
&\simeq&  (\delta^{\ell}_{31})_{LL}
\simeq -1.7 \times 10^{-5} ~\Delta a_L^{\ell}, \nn\\
(\delta^{\ell}_{23})_{LL}
&\simeq& (\delta^{\ell}_{32})_{LL}
\simeq  8.4 \times 10^{-8} ~\Delta a_L^{\ell},\nn\\
(\delta^{\ell}_{12})_{RR}
&\simeq& (\delta^{\ell}_{21})_{RR}
\simeq 8.4 \times 10^{-8} ~\Delta a_R^{\ell}, \label{deltaell}
\\
(\delta^{\ell}_{13})_{RR}
&\simeq& (\delta^{\ell}_{31})_{RR}
\simeq 5.9 \times 10^{-2} ~\Delta a_R^{\ell}, \nn\\
(\delta^{\ell}_{23})_{RR}
&\simeq& (\delta^{\ell}_{32})_{RR} 
\simeq -1.4 \times 10^{-6} ~\Delta a_R^{\ell}.\nn
\ee

\vskip 0.5cm
\underline{\bf Leptonic sector ($LR$): }
\be
Im (\delta^{\ell}_{ij})_{LR} &=&0,\nn\\
(\delta^{\ell}_{12})_{LR}
&\simeq& 5.1 \times 10^{-6} ~
\left(\tilde{A}_2^{\ell}-\tilde{A}_4^{\ell}\right)
\left(\frac{100~\mbox{GeV}}{m_{\tilde{\ell}} }\right), \nn\\
(\delta^{\ell}_{21})_{LR}
&\simeq& 2.5 \times 10^{-8} ~
\left(\tilde{A}_2^{\ell}-\tilde{A}_4^{\ell}\right)
\left(\frac{100~\mbox{GeV}}{m_{\tilde{\ell}} }\right), 
\label{LRell}\\
(\delta^{\ell}_{13})_{LR}
&\simeq& 3.1 \times 10^{-7} ~
\left(\tilde{A}_4^{\ell}-\tilde{A}_5^{\ell}\right)
\left( \frac{100~\mbox{GeV}}{m_{\tilde{\ell}} }\right), \nn\\
(\delta^{\ell}_{31})_{LR}
&\simeq& 1.1 \times 10^{-3} ~
\left(\tilde{A}_2^{\ell}-\tilde{A}_5^{\ell}\right)
\left( \frac{100~\mbox{GeV}}{m_{\tilde{\ell}} }\right), \nn\\
(\delta^{\ell}_{23})_{LR}
&\simeq&  -1.5 \times 10^{-9} ~
\left(\tilde{A}_4^{\ell}-\tilde{A}_5^{\ell}\right)
\left( \frac{100~\mbox{GeV}}{m_{\tilde{\ell}} }\right),\nn\\
(\delta^{\ell}_{32})_{LR}
&\simeq& -2.5 \times 10^{-8} ~
\left(\tilde{A}_2^{\ell}-\tilde{A}_5^{\ell}\right)
\left( \frac{100~\mbox{GeV}}{m_{\tilde{\ell}} }\right).\nn
\ee

\vskip 0.5cm
\underline{\bf Up quark  sector ($LL$ and $RR$):}
\be
(\delta^u_{12})_{LL}
& \simeq&  (\delta^d_{21})_{LL}\simeq 
-1.1 \times 10^{-4}~\Delta a_L^{u}, \nn\\
(\delta^u_{13})_{LL}
& \simeq& (\delta^d_{31})_{LL}
\simeq  -3.2\times 10^{-3}~\Delta a_L, \nn\\
(\delta^u_{23})_{LL}
& \simeq& (\delta^d_{32})_{LL}
\simeq  -3.3 \times 10^{-2}~\Delta a_L^{u},\nn\\
(\delta^u_{12})_{RR}
& \simeq& (\delta^u_{21})_{RR}
\simeq 7.2 \times 10^{-4} ~\Delta a_R^{u}, \label{deltau}
\\
(\delta^u_{13})_{RR}
& \simeq& (\delta^u_{31})_{RR}
\simeq - 8.2 \times 10^{-3}~\Delta a_R^{u}, \nn\\
(\delta^u_{23})_{RR}
& \simeq& (\delta^u_{32})_{RR}
\simeq 8.6 \times 10^{-2}~\Delta a_R^{u}.\nn
\ee

\vskip 0.5cm
\underline{\bf Up quark  sector  ($LR$)}:
\be
(\delta^{u}_{11})_{LR}
&\simeq& 
\left[ -1.3\times 10^{-4} \tilde{A}_{1}^{u}
 +1.4\times 10^{-4} \tilde{A}_{2}^{u}
~\right] 
\left( \frac{500 ~\mbox{GeV}}{m_{\tilde{q}} }\right), \nn\\
(\delta^{u}_{12})_{LR}
&\simeq&-(\delta^{u}_{21})_{LR}\simeq 
-9.7 \times 10^{-5} \left(\tilde{A}_{2}^{u}+\tilde{A}_{3}^{u}-
\tilde{A}_{4}^{u}-\tilde{A}_{5}^{u}
\right)~
\left( \frac{500 ~\mbox{GeV}}{m_{\tilde{q}} }\right) ,\nn\\
(\delta^{u}_{13})_{LR}
&\simeq& 
1.1 \times 10^{-3} \left(\tilde{A}_{3}^{u}-\tilde{A}_{5}^{u}
\right)~
\left( \frac{500 ~\mbox{GeV}}{m_{\tilde{q}} }\right), \nn\\
(\delta^{u}_{31})_{LR}
&\simeq& 
2.9 \times 10^{-3} \left(\tilde{A}_{3}^{u}-\tilde{A}_{4}^{u}
\right)~
\left( \frac{500 ~\mbox{GeV}}{m_{\tilde{q}} }\right),
\label{LRu}\\
(\delta^{u}_{23})_{LR}
&\simeq&1.2 \times 10^{-2} \left(\tilde{A}_{3}^{u}
-\tilde{A}_{5}^{u}
\right)~
\left( \frac{500 ~\mbox{GeV}}{m_{\tilde{q}} }\right) ,\nn\\
(\delta^{u}_{32})_{LR}
&\simeq&- 3.0 \times 10^{-2} \left(\tilde{A}_{3}^{u}-\tilde{A}_{4}^{u}
\right)~
\left( \frac{500 ~\mbox{GeV}}{m_{\tilde{q}} }\right).\nn
\ee

\vskip 0.5cm
\underline{\bf Down quark  sector ($LL$ and $RR$):}
\be
(\delta^d_{12})_{LL}
& \simeq&  (-6.1 \times 10^{-4}-3.8 \times 10^{-4}~I) ~\Delta a_L^{d}, \nn\\
(\delta^d_{21})_{LL}
& \simeq&  (-6.1 \times 10^{-4}+3.8 \times 10^{-4}~I) ~\Delta a_L^{d}, \nn\\
(\delta^d_{13})_{LL}
& \simeq&  (1.1\times 10^{-2}+1.5 \times 10^{-4}~I) ~\Delta a_L^{d}, \nn\\
(\delta^d_{31})_{LL}
& \simeq&  (1.1\times 10^{-2}-1.5 \times 10^{-4}~I) ~\Delta a_L^{d}, \nn\\
(\delta^d_{23})_{LL}
& \simeq&  (5.3 \times 10^{-2}- 3.2 \times 10^{-2}~I) ~\Delta a_L^{d},\nn\\
(\delta^d_{32})_{LL}
& \simeq&  (5.3 \times 10^{-2}+3.2 \times 10^{-2}~I) ~\Delta a_L^{d},\label{deltad}
\\
(\delta^d_{12})_{RR}
& \simeq& (-3.7 \times 10^{-2} + 2.0 \times 10^{-2}~I) ~\Delta a_R^{d}, \nn\\
(\delta^d_{21})_{RR}
& \simeq& (-3.7 \times 10^{-2} - 2.0 \times 10^{-2}~I) ~\Delta a_R^{d}, \nn\\
(\delta^d_{13})_{RR}
& \simeq& (- 2.6 \times 10^{-2} + 6.9 \times 10^{-2}~I) ~\Delta a_R^{d}, \nn\\
(\delta^d_{31})_{RR}
& \simeq& (- 2.6 \times 10^{-2} -6.9 \times 10^{-2}~I) ~\Delta a_R^{d}, \nn\\
(\delta^d_{23})_{RR}
& \simeq& (-3.2 \times 10^{-1} +2.9 \times 10^{-1}~I) ~\Delta a_R^{d},\nn\\
(\delta^d_{32})_{RR}
& \simeq& (-3.2 \times 10^{-1} -2.9 \times 10^{-1}~I) ~\Delta a_R^{d}.\nn
\ee

\vskip 0.5cm
\underline{\bf Down quark  sector ($LR$): }
\be
(\delta^{d}_{11})_{LR}
&\simeq& 
\left[ -1.6 \tilde{A}_{1}^{d}+2.3 \tilde{A}_{2}^{d}
-I~0.6(\tilde{A}_{1}^{d}-\tilde{A}_{2}^{u})
 +I~0.5(\tilde{A}_{3}^{d}+\tilde{A}_{2}^{d}-\tilde{A}_{4}^{d}-\tilde{A}_{5}^{d})
~\right] \nn\\
 & &\times 10^{-5}~\left( \frac{500 ~\mbox{GeV}}{m_{\tilde{q}} }\right), \nn\\
 (\delta^{d}_{12})_{LR}
&\simeq&\left[~
(2.4+1.7 ~I)\times 10^{-5} \left(\tilde{A}_{2}^{d}-\tilde{A}_{5}^{d}
\right)+
(2.2+1.9 ~I)\times 10^{-5} \left(\tilde{A}_{3}^{d}-\tilde{A}_{4}^{d}
\right)\right]\nn\\
& &\times\left( \frac{500 ~\mbox{GeV}}{m_{\tilde{q}} }\right), \nn\\
(\delta^{d}_{21})_{LR}
&\simeq&\left[~
(2.7+1.9 ~I)\times 10^{-5} \left(\tilde{A}_{2}^{d}-\tilde{A}_{5}^{d}
\right)+
(2.0+1.7 ~I)\times 10^{-5} \left(\tilde{A}_{3}^{d}-\tilde{A}_{4}^{d}
\right)\right]\nn\\
& &\times\left( \frac{500 ~\mbox{GeV}}{m_{\tilde{q}} }\right),
\label{LRd}
\ee
\be
(\delta^{d}_{13})_{LR}
&\simeq& -\left[~
(4.1+2.9~I) \times 10^{-5} \left(\tilde{A}_{3}^{d}-\tilde{A}_{5}^{d}
\right)+
(1.4+0.1~I) \times 10^{-5}
\left(\tilde{A}_{2}^{d}-\tilde{A}_{4}^{d}
\right) ~\right]\nn\\
& &
\times\left( \frac{500 ~\mbox{GeV}}{m_{\tilde{q}} }\right), \nn\\
(\delta^{d}_{31})_{LR}
&\simeq& 
-(3.4+2.4~ I) \times 10^{-4} \left(\tilde{A}_{3}^{d}-\tilde{A}_{4}^{d}
\right)~
\left( \frac{500 ~\mbox{GeV}}{m_{\tilde{q}} }\right) ,\nn\\
(\delta^{d}_{23})_{LR}
&\simeq&\left[~
-2.6 \times 10^{-4} \left(\tilde{A}_{3}^{d}-\tilde{A}_{4}^{d}
\right)
+3.4 \times 10^{-4} \left(\tilde{A}_{4}^{d}-\tilde{A}_{5}^{d}
\right)\right. \nn\\
& &\left.+I~\left(\tilde{A}_{1}^{d}+\tilde{A}_{2}^{d}
-3\tilde{A}_{3}^{d}-\tilde{A}_{4}^{d}
+2 \tilde{A}_{5}^{d}
\right)\times 10^{-5}~\right]
\left( \frac{500 ~\mbox{GeV}}{m_{\tilde{q}} }\right), \nn\\
(\delta^{d}_{32})_{LR}
&\simeq& (2.5 +0.2~I)\times 10^{-3}
 \left(\tilde{A}_{3}^{d}-\tilde{A}_{4}^{d}
\right)~
\left( \frac{500 ~\mbox{GeV}}{m_{\tilde{q}} }\right).\nn
\ee
Comparing the $\delta$'s above with Table I, we see
that the experimental bounds for the most of the cases are satisfied,
 if $|\Delta a|$'s  and  $|(\tilde{A}_{i}-\tilde{A}_{j})|$'s are
less than about one. 
The experimental constraints coming from
the CP violations in the $K^{0}-\bar{K}^{0}$ system  on $\delta^{d}_{12}$,
more precisely on $\sqrt{|\mbox{Im}(\delta^d_{12})^2_{LL,RR}|},
\sqrt{|\mbox{Im}(\delta^d_{12})_{LL}(\delta^d_{12})_{RR}|}$
and $|\mbox{Im}(\delta^d_{12})_{LR}|$
are very severe.
Note however, one of the most strong constraint coming
from $\epsilon'/\epsilon$ on
$|\mbox{Im}(\delta_{12}^{d})_{LR}|$ is  satisfied.
It seems at first sight that
$(\delta^d_{12})_{RR}$ is too large
to satisfy $\sqrt{|\mbox{Im}(\delta^d_{12})^2_{RR}|} < 3.2 \times 10^{-3}
\tilde{m}_{\tilde{q}}$, because $(\delta^d_{12})_{RR}\sim O(10^{-2})
\Delta a^{d}_{R}$ in the present model 
as we can see from (\ref{deltad}).
Fortunately, this is not the case.
First of all, all the $a$'s are real, because the soft scalar mass matrices are
diagonal by $S_{3}$ symmetry.
Secondly, there is a freedom to make a phase rotation
of the fermion mass eigenstates without changing 
their masses and $V_{\rm CKM}$.
Under this phase rotation, the unitary matrices undergo as
\be
U_{dL,R} &\to & U_{dL,R} \times 
 \left(
\begin{array}{ccc}
e^{i \varphi_{1}} & 0 & 0\\
0 &e^{i \varphi_{2}} & 0\\
0 & 0 & e^{i \varphi_{3}}
\end{array}\right).
\ee
Then $\delta^{d}_{ij}$ change
to
\be
\delta^{d}_{ij}\times \exp -i( \varphi_{i}-\varphi_{j}),
\label{phaserotation}
\ee
which means
\be
(\delta^d_{12})_{RR} &\to &
(-3.7 \times 10^{-2} + 2.0 \times 10^{-2}~I) 
~\exp -i( \varphi_{1}-\varphi_{2})~\Delta a_R^{d}.
\ee
Therefore, it is always possible to make $(\delta^d_{12})_{RR}$
and simultaneously $(\delta^d_{21})_{RR}$
real so that the constraints on  $|\mbox{Im}(\delta^d_{12})^2_{RR}|$
and  $|\mbox{Im}(\delta^d_{12})_{LL}(\delta^d_{12})_{RR}|$ can be satisfied.
Note that  the phase rotation (\ref{phaserotation})
has practically no influence on the other constraints.
The constraints from
the electric dipole moment (EDM) of the neutron
on $|\mbox{Im}(\delta^d_{11})_{LR}|$
 and $|\mbox{Im}(\delta^u_{11})_{LR}|$
 are also very severe\footnote{
 The EDM of the electron in the present model is practically zero, because
all the Yukawa couplings $Y^{E}$ and also the 
corresponding trilinear couplings $h^{E}$
can be made real by appropriate phase rotations
of the corresponding fermions and their scalar partners.
A possible source for the EDM of the electron 
is the complex Yukawa couplings $Y^{N}_{3}$ and $Y^{N}_{4}$
in (\ref{potn}).}.
 It is not possible to make them real by the phase rotation (\ref{phaserotation}).
 Therefore, we have to fine tune certain $A$'s.
Note, however, that  it is possible to make
 $A_{2}^{q},A_{4}^{q}$ and $A_{5}^{q}$ real 
 by an appropriate phase rotation
of the  squarks. In this case we have to assume that
\be
|\mbox{Im} (\tilde{A}_{1}^{u})| &\lsim &O(10^{-2}) ~~ \mbox{and}~~
|\mbox{Im} (\tilde{A}_{1,3}^{d})| \lsim O(10^{-1}),
\label{A13}
\ee
while their real parts can be $O(1)$.

From the analyses in this section,
we conclude that,
apart from the fine tuning (\ref{A13}), 
the FCNCs and CP phases, which are induced by
the SSB parameters in $O(1)$ disorder at  $M_{\rm SUSY}$,
are  sufficiently suppressed to satisfy
the experimental constraints.
This is a consequence of  $S_{3}$ flavor  symmetry.

\begin{table}[thb]
\begin{center}
\begin{tabular}{|c|c||c|c|} 
 \hline 
 &  Exp. bound  & 
  & Exp. bound 
   \\ \hline 
$|(\delta^{\ell}_{12})_{LL,RR}|$ &
 $7.7 \times 10^{-3} ~\tilde{m}^2_{\tilde{\ell}}$ 
  & $|(\delta^{\ell}_{12})_{LR}|$ &
   $ 1.7 \times 10^{-6} ~\tilde{m}^2_{\tilde{\ell}} $
\\ \hline
$|(\delta^{\ell}_{13})_{LL,RR}|$ & 
$ 29 ~ \tilde{m}^2_{\tilde{\ell}}$ &
$|(\delta^{\ell}_{13})_{LR}|$
& $ 1.1 \times 10^{-1} ~ \tilde{m}^2_{\tilde{\ell}} $
\\ \hline
$|(\delta^{\ell}_{23})_{LL,RR}|$ & 
$ 5.3  ~ \tilde{m}^2_{\tilde{\ell}} $
& $|(\delta^{\ell}_{23})_{LR}|$
& $ 2.0 \times 10^{-2}  ~ \tilde{m}^2_{\tilde{\ell}} $
\\ \hline
$\sqrt{|\mbox{Re}(\delta^d_{12})^2_{LL,RR}|}$
& $4.0 \times 10^{-2} ~\tilde{m}_{\tilde{q}}$ 
& $\sqrt{|\mbox{Re}(\delta^d_{12})_{LL}(\delta^d_{12})_{RR}|}$
& $2.8 \times 10^{-3} ~\tilde{m}_{\tilde{q}}$
\\ \hline
$\sqrt{|\mbox{Re}(\delta^d_{12})^2_{LR}|}$
 & $4.4 \times 10^{-3} ~\tilde{m}_{\tilde{q}}$
& $\sqrt{|\mbox{Re}(\delta^d_{13})^2_{LL,RR}|}$
 & $9.8 \times 10^{-2} ~\tilde{m}_{\tilde{q}}$ 
\\ \hline
$\sqrt{|\mbox{Re}(\delta^d_{13})_{LL}(\delta^d_{13})_{RR}|}$
&  $1.8 \times 10^{-2} ~\tilde{m}_{\tilde{q}}$
& $\sqrt{|\mbox{Re}(\delta^d_{13})^2_{LR}|}$ 
& $3.3 \times 10^{-3} ~\tilde{m}_{\tilde{q}}$
\\ \hline 
$\sqrt{|\mbox{Re}(\delta^u_{12})^2_{LL,RR}|}$ 
& $1.0 \times 10^{-1} ~\tilde{m}_{\tilde{q}}$
& $\sqrt{|\mbox{Re}(\delta^u_{12})_{LL}(\delta^u_{12})_{RR}|}$
& $1.7 \times 10^{-2} ~\tilde{m}_{\tilde{q}}$
\\ \hline
$\sqrt{|\mbox{Re}(\delta^u_{12})^2_{LR}|}$
& $3.1 \times 10^{-3} ~\tilde{m}_{\tilde{q}}$
 & $\sqrt{|\mbox{Im}(\delta^d_{12})^2_{LL,RR}|}$
 & $3.2 \times 10^{-3}~ \tilde{m}_{\tilde{q}}$
 \\ \hline
 $\sqrt{|\mbox{Im}(\delta^d_{12})_{LL}(\delta^d_{12})_{RR}|}$
& $2.2 \times 10^{-4} ~\tilde{m}_{\tilde{q}}$
& $\sqrt{|\mbox{Im}(\delta^d_{12})^2_{LR}|}$
& $3.5 \times 10^{-4} ~\tilde{m}_{\tilde{q}}$
 \\ \hline
$|(\delta^d_{23})_{LL,RR}|$
 & $8.2~ \tilde{m}_{\tilde{q}}^2$
& $|(\delta^d_{23})_{LR}|$
& $1.6 \times 10^{-2} ~\tilde{m}_{\tilde{q}}^2$
\\ \hline 
$|\mbox{Im}(\delta^d_{12})_{LL,RR}|$
 & $4.8 \times 10^{-1} ~\tilde{m}_{\tilde{q}}^2$
& $|\mbox{Im}(\delta^d_{12})_{LR}|$
& $2.0 \times 10^{-5} ~\tilde{m}_{\tilde{q}}^2$
\\ \hline 
$|\mbox{Im}(\delta^d_{11})_{LR}|$
 & $3.0 \times 10^{-6} ~\tilde{m}_{\tilde{q}}^2$
& $|\mbox{Im}(\delta^u_{11})_{LR}|$
& $5.9 \times 10^{-6} ~\tilde{m}_{\tilde{q}}^2$
 \\ \hline
$|\mbox{Im}(\delta^\ell_{11})_{LR}|$
 & $3.7\times 10^{-7}~ \tilde{m}_{\tilde{\ell}}^2$
& & 
\\ \hline
\end{tabular}
\caption{Experimental bounds on  $\delta$'s, where
the parameters $\tilde{m}_{\tilde{\ell}}$
and  $\tilde{m}_{\tilde{q}}$ denote
$m_{\tilde{\ell}} /100$ GeV and ~~~~~~~~~~~
$m_{\tilde{\ell}} /500$ GeV,
respectively.
See \cite{fcnc} for details.}
\end{center}
\end{table}

\section{
Suppression of FCNCs at the compactification scale}
As we have seen in the previous section, $S_{3}$ flavor symmetry can
 suppress FCNCs and  dangerous CP violating phases
at $M_{\rm SUSY}$.
There, we have assumed that
$\Delta a$'s and $\tilde{A}$'s are $O(1)$
at $M_{\rm SUSY}$ (they are defined in (\ref{deltaAt1})).
At a more fundamental scale,
they may be  $O(10)$ or $O(100)$, for instance.
Then we need another suppression mechanism
to bring down say, an  $O(10)$ disorder 
to  an $O(1)$ disorder at $M_{\rm SUSY}$.
In this section we would like to discuss the second stage
of  suppressing FCNCs and CP phases.
As shown in \cite{kubo6,choi1,choi2,kaji1}, 
the IR attractive force of gauge interactions can be
amplified in extra dimensions, driving 
the SSB terms to their infrared attractive values at the
compactification scale
$\Lambda_{C}$.
If the disordered SSB terms at 
the cutoff scale $\Lambda$
converge rapidly to their
IR fixed values that are flavor blind,
  FCNCs  and dangerous CP violating phases can be 
desirably  suppressed.

\subsection{A model }
Let us consider embedding the supersymmetric extension 
of the $S_3$ invariant SM of Sect.  II  into $\delta$ extra dimensions 
compactified on an orbifold with a compactification scale 
$\Lambda_{C}=1/R$.
We assume that the matter multiplets and the $U(1)_Y$ 
gauge multiplet are located 
at a fixed point, while the $SU(2)_L$ and $SU(3)_C$ 
gauge multiplets propagate
in the bulk. Therefore, 
only the KK tower of the $SU(2)_L$ and $SU(3)_C$ gauge
multiplets 
contributes to the power running of parameters.
Thanks to the dominance of the power running,
we may ignore the logarithmic contributions.
We, however,  include
those logarithmic corrections that come from $g_1$.
Under this approximation we have computed 
the $\beta$ functions for the rigid and soft breaking parameters,
which are given in Appendix.

First we recall that the KK tower of the $U(1)_{Y}$ gauge multiplet would 
not contribute to the $\beta$ function of $g_{1}$ and hence that of $M_{1}$
(at the one-loop level).
So, they would be basically constant.
However, they would contribute to the $\beta$ functions of 
the Yukawa couplings and hence to those of the SSB parameters.
Since $g_{1}$ and $M_{1}$ remain approximately constant (while
the other couplings become smaller) as  $\Lambda$ increases,
the contributions of the  $U(1)_{Y}$ KK modes
to the RG evolution would become dominant at high energies.
As a consequence, 
we obtain a power dependence of $\Lambda$ 
instead of the logarithmic dependence 
(which is hidden in $\varepsilon_{2,3}$
of the solutions  (\ref{solution-quark}), 
(\ref{solution-lepton}) (\ref{solution-higgs})). 
In other words, the  uncertainty at high energy would remain
unsuppressed at low energy,
which would not be  desirable.

Next let us  discuss why we include the contributions from $g_{1} $.
The discussion will also clarify the restriction of the 
suppression mechanism of the model.
As we see from (\ref{power-gauge1})
and (\ref{power-gaugino1}), the $U(1)_{Y}$ gauge coupling $g_{1}$ and 
the corresponding gaugino mass $M_{1}$
are running only logarithmically  above
 $\Lambda_{C}$. 
 So they are basically constant
 compared with other power-law-running parameters.
 To see this more in detail, we recall the analytic solutions
 for the RG evolution equations
\be
\frac{1}{g_1^2(\Lambda)}&=&\frac{1}{g_1^2(\Lambda_C)}-\frac{2(\sum
l(R))}{16\pi^2}\log \frac{\Lambda}{\Lambda_C},\label{sol_abelian} \\
\frac{1}{g^2_2(\Lambda)}&=&
\frac{1}{g^2_2(\Lambda_C)}+\frac{C_2(SU(2))X_\delta}{8\pi^2\delta}\left\{
\left(\frac{\Lambda}{\Lambda_C} \right)^\delta-1
\right\},\label{sol_nonabelian1} \\
\frac{1}{g^2_3(\Lambda)}&=&
\frac{1}{g^2_3(\Lambda_C)}+\frac{C_2(SU(3))X_\delta}{8\pi^2\delta}\left\{
\left(\frac{\Lambda}{\Lambda_C} \right)^\delta-1
\right\},
\label{sol_nonabelian2} 
\ee
where $(\sum l(R))=39/5, C_2(SU(N))=N$.
For $\Lambda/\Lambda_C >>1 $, 
Eqs.~(\ref{sol_nonabelian1}) and (\ref{sol_nonabelian2}) may be further
approximated to
\be
\left[\frac{g_a(\Lambda_C)}{g_a(\Lambda)}\right]^{2} &\simeq &
\bigg[\frac{C_2(G_a)X_\delta g_{a}^2(\Lambda_C)}{8\pi^2\delta}\bigg]
\bigg(\frac{\Lambda}{\Lambda_C}\bigg)^{\delta} ~( a=2,3)\nn\\
&\simeq & 
\left\{ \begin{array}{c}
(0.02~ \Lambda/\Lambda_C)~\mbox{for} ~a=2
\\ 
(0.1 ~\Lambda/\Lambda_C)~\mbox{for} ~a=3
\end{array}\right.~\mbox{with}~\delta=1\nn\\
&\simeq & 
\left\{ \begin{array}{c}
(0.017~ \Lambda/\Lambda_C)^{2}~\mbox{for} ~a=2
\\ 
(0.09 ~\Lambda/\Lambda_C)^{2}~\mbox{for} ~a=3
\end{array}\right.~\mbox{with}~\delta=2,
\label{enhancement}
\ee
where  $X_\delta=\pi^{\delta/2}/\Gamma(1+\delta/2)=2(\pi)$ for $\delta=1(2)$
has been used, and we have assumed
that $g_{2}^{2}(M_{Z})\simeq g_{2}^{2}(\Lambda_{C})\simeq  0.034 
\times 4 \pi$ and 
$g_{3}^{2}(M_{Z})\simeq g_{3}^{2}(\Lambda_{C})\simeq  0.12 \times 4 \pi$.
For the gaugino masses we obtain
 \be
M_a(\Lambda_C)&=&\bigg(\frac{g_a(\Lambda_C)}{g_a(\Lambda)}
\bigg)^2M_a(\Lambda),
\qquad a=1,2,3.
 \label{gauge-gaugino}
\ee
Note that $M_{1}$
 does not get a power enhancement
in contrast to $M_{2,3}$,
because $(g_{1}(\Lambda_{C})/g_{1}(\Lambda))^2$
is only logarithmic. Therefore, $M_{1}$ should be much larger
than $M_{2,3}$  at the cutoff scale $\Lambda$, if  $M_{1}(\Lambda_{C})$
should be $O(M_{2,3}(\Lambda_{C}))$.
The logarithmic running of $g_{1}$ and $M_{1}$ has an important consequence on 
the soft scalar mass matrix  
for the right-handed sleptons $m^{2}_{E_{R}}$.
As we see from (\ref{betamER}), 
their  $\beta$ functions depend
only on $g_{1}$ and $M_{1}$, and one finds that
\begin{eqnarray}
(m^2_{E_R})^i_j(\Lambda_C)&=(m^2_{E_R})^i_j(\Lambda)
+\delta^i_j\frac{24}{5}\varepsilon_3|M_1|^2(\Lambda_C),
\label{mER}
\end{eqnarray}
where $i,j=1,2,3$ and 
\be
\varepsilon_n&=&\frac{g_1^2(\Lambda_C)}{16\pi^2}
\int_{\Lambda_C}^{\Lambda}\frac{d\Lambda}{\Lambda}
\bigg(\frac{g_1(\Lambda_C)}{g_1(\Lambda)}\bigg)^{2n}\nonumber\\
& = & \frac{g_1^2(\Lambda_C)}{16\pi^2}
\ln (\Lambda/\Lambda_{C}) +O(g_{1}^{4}).
\label{epsilon}
\ee
So, $  m_{E_{R}}^{2}$, too,
get only  logarithmic corrections, which means that
$m_{E_{R}}^{2}(\Lambda_{C}) \sim m_{E_{R}}^{2}(\Lambda)$.
Therefore, $m_{E_{R}}^{2}(\Lambda)$ should be much larger
than $M_{2,3}^{2}(\Lambda)$;
too small  $m_{E_{R}}^{2}(\Lambda_{C})$ are phenomenologically not
acceptable (in contrast to $M_{1}(\Lambda_{C}$)).
However, if $m_{E_{R}}^{2}(\Lambda)>> M_{2}^{2}(\Lambda)$, 
say $m_{E_{R}}^{2}(\Lambda)\sim(10^{3} M_{2}(\Lambda))^{2}$,
this  would be a unnatural fine tuning  at the cutoff scale
$\Lambda$, and would be 
against our philosophy of the present paper.
So, we require 
\be
|m^2_{E_R}|(\Lambda)& \lsim
(10 |M_2|(\Lambda))^{2}
\label{restriction}
\ee
for the fine tuning not to be  unnatural.
This is a strong restriction of the suppression mechanism
for the present model, on one hand,
because it limits $M_{2}(\Lambda_{C})/M_{2}(\Lambda)$
and hence $\Lambda/\Lambda_{C}$
which is the power-law enhancement factor 
 (\ref{enhancement}).
On the other hand, it is a prediction 
of the model that $M_{1}(\Lambda_{C})$ and 
$m_{E_{R}}(\Lambda_{C})$ are smaller than
the other gaugino masses and soft scalar masses.
Further, as we see from (\ref{mER}), 
$(m^2_{E_R})^i_j$ do not converge to
IR attractive values. Fortunately, 
as we can see from Table I and 
(\ref{deltaell}), 
the experimental bound on $\delta_{RR}$'s
are very weak; it is sufficient to satisfy
\be
\Delta a_{R}^{\ell} &< & 5\times 10^{2} 
(\frac{m_{\ell }}{100~\mbox{GeV}})^{2}.
\ee

\subsection{Suppression of FCNCs}
Keeping the above discussions in mind, we proceed
with our consideration of the IR attractiveness
of the other SSB parameters.
Using the $\beta$ functions given in Appendix, we find the following
analytic solutions for the RG evolution equations.

\vskip 0.5cm 
\underline{\bf Quark sector:}
\vskip 0.5cm
\dis{
\frac{h^D_k}{Y^D_k}(\Lambda_C)&=\frac{h^D_k}{Y^D_k}(\Lambda)-\frac{14}{15}\varepsilon_2M_1(\Lambda_C) -\frac{12}{8}(1-c)
M_2(\Lambda_C)-\frac{16}{9}(1-d)M_3(\Lambda_C),\\
\frac{h^U_k}{Y^U_k}(\Lambda_C)&=\frac{h^U_k}{Y^U_k}(\Lambda)-\frac{26}{15}\varepsilon_2M_1(\Lambda_C) -\frac{12}{8}(1-c)M_2(\Lambda_C)
-\frac{16}{9}(1-d)M_3(\Lambda_C),\label{solution-quark}\\
(m^2_Q)^i_j(\Lambda_C)&=(m^2_Q)^i_j(\Lambda)
+\delta^i_j\bigg[\frac{2}{15}\varepsilon_3|M_1|^2(\Lambda_C)
+\frac{6}{8}(1-c^2)|M_2|^2(\Lambda_C)
+\frac{32}{36}(1-d^2)|M_3|^2(\Lambda_C)\bigg],\\
(m^2_{U_R})^i_j(\Lambda_C)&=(m^2_{U_R})^i_j(\Lambda)
+\delta^i_j\bigg[\frac{32}{15}\varepsilon_3|M_1|^2(\Lambda_C)
+\frac{32}{36}(1-d^2)|M_3|^2(\Lambda_C)\bigg],\\
(m^2_{D_R})^i_j(\Lambda_C)&=(m^2_{D_R})^i_j(\Lambda)
+\delta^i_j\bigg[\frac{8}{15}\varepsilon_3|M_1|^2(\Lambda_C)
+\frac{32}{36}(1-d^2)|M_3|^2(\Lambda_C)\bigg],
} 
where $k=1,2,3,4,5$, $i,j=1,2,3$,
$\varepsilon_{n}$ is defined in (\ref{epsilon}), and 
\be
c&=&\bigg(\frac{g_2(\Lambda)}{g_2(\Lambda_C)}\bigg)^2,~
d=\bigg(\frac{g_3(\Lambda)}{g_3(\Lambda_C)}\bigg)^2.
\ee

\vskip 0.5cm 
\underline{\bf Lepton sector:}
\vskip 0.5cm
\dis{
\frac{h^E_k}{Y^E_k}(\Lambda_C)&=\frac{h^E_k}{Y^E_k}(\Lambda)-\frac{18}{5}\varepsilon_2 M_1(\Lambda_C)-\frac{12}{8}(1-c)M_2(\Lambda_C),\\
\frac{h^N_k}{Y^N_k}(\Lambda_C)&=\frac{h^N_k}{Y^N_k}(\Lambda)-\frac{6}{5}\varepsilon_2 M_1(\Lambda_C)-\frac{12}{8}(1-c)M_2(\Lambda_C),\\
(m^2_L)^i_j(\Lambda_C)&=(m^2_L)^i_j(\Lambda)
+\delta^i_j\bigg[\frac{6}{5}\varepsilon_3|M_1|^2(\Lambda_C)
+\frac{6}{8}(1-c^2)|M_2|^2(\Lambda_C)\bigg],\label{solution-lepton}\\
(m^2_{E_R})^i_j(\Lambda_C)&=(m^2_{E_R})^i_j(\Lambda)
+\delta^i_j\frac{24}{5}\varepsilon_3|M_1|^2(\Lambda_C),
}
where $k=2,4,5(2,3,4)$ for the charged lepton(neutrino) sector and $i,j=1,2,3$.
The Majorana neutrino masses, $M_{1,3}$, and their soft scalar masses, 
$(m^2_{N_R})^i_j$, have only a logarithmic running. 

\vskip 0.5cm
\underline{\bf Higgs sector:}
\vskip 0.5cm
\dis{
\frac{B_k}{\mu_k}(\Lambda_C)&=\frac{B_k}{\mu_k}(\Lambda)-\frac{6}{5}\varepsilon_2M_1(\Lambda_C)-
\frac{12}{8}(1-c)M_2(\Lambda_C), \ \ k=1,3, \label{solution-higgs}\\
m^2_a(\Lambda_C)&=m^2_a(\Lambda)+\frac{6}{5}\varepsilon_3|M_1|^2(\Lambda_C) 
+\frac{6}{8}(1-c^2)|M_2|^2(\Lambda_C), \ \ a=H^U,H^D.
}
As it is clear,   the quantities on the right-hand side
of (\ref{solution-quark}), (\ref{solution-lepton})
 and (\ref{solution-higgs}) whose argument
is  $\Lambda$ are initial values at $\Lambda$,
 and those of $\Lambda_{C}$ are
 IR attractive values. Therefore, $\Delta a$'s and also
 $\tilde{A}$'s, given in  (\ref{deltaAt1}),  at $\Lambda_{C}$ can be written 
 in terms of the initial values.
 We emphasize that all the infrared attractive values are flavor diagonal.
 
 \begin{figure}[tb]
\label{fig:s3-figure-trilinear.eps}
\includegraphics*[width=0.6\textwidth]{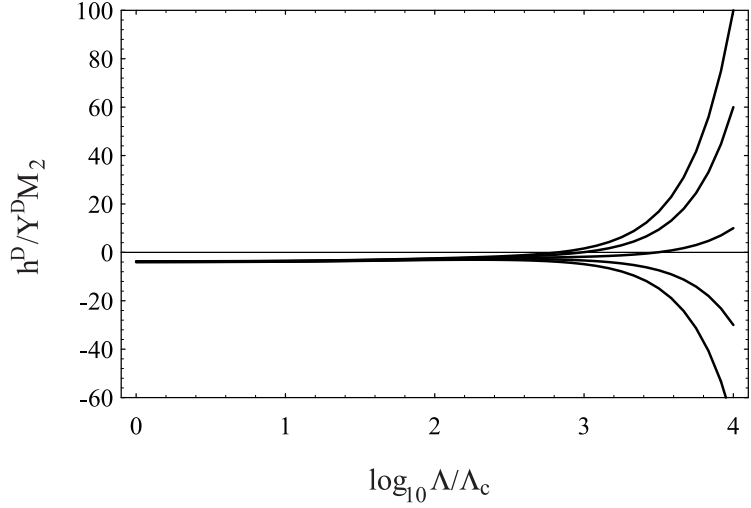}
\caption{The IR attractiveness of  $h_D/(M_2Y_D)$.}
\end{figure}

\begin{figure}[tb]
\label{fig:s3-figure-softmass.eps}
\includegraphics*[width=0.6\textwidth]{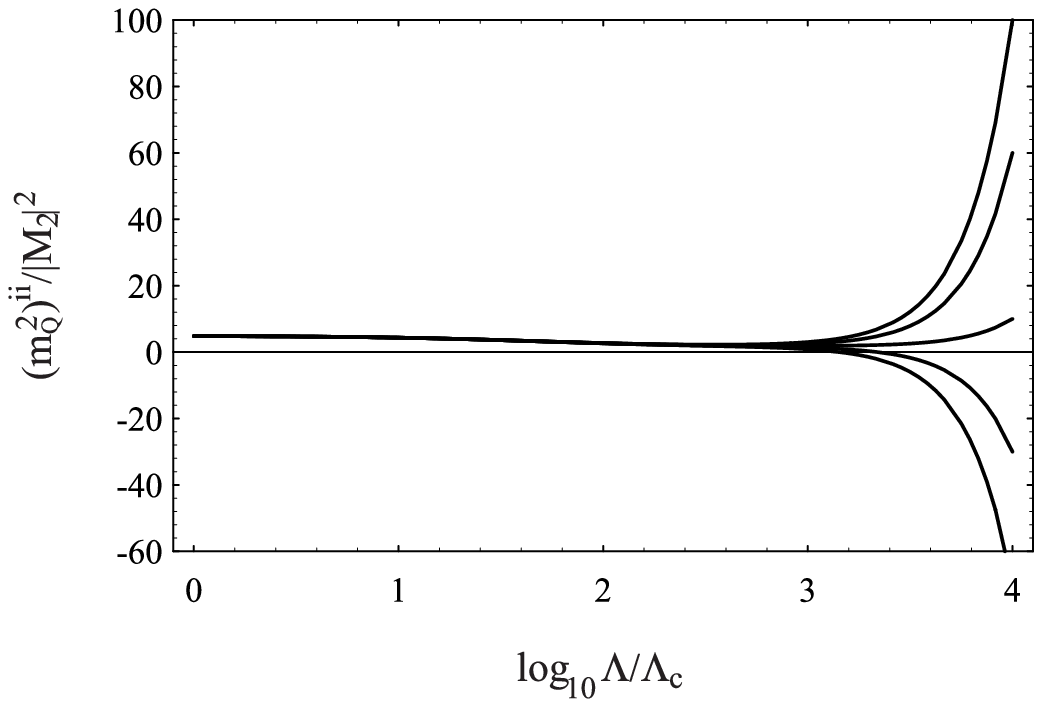}
\caption{The IR attractiveness of $m^2_Q/|M_2|^2$.}
\end{figure}

To estimate the suppression of the
initial values at $\Lambda$, we assume that the initial SSB parameters are disordered
around  $M_{2}(\Lambda)$. For instance,
 \be
 A_{1}^{U}(\Lambda) &=&
 \frac{h_{1}^U}{Y_{1}^{U}}(\Lambda)
= \kappa_{1}^{U} M_{2}(\Lambda),
~m^{2}_{Q} =
 \kappa_{m_{Q}}^{2} M_{2}^{2}(\Lambda),
 \ee
and  similarly for other SSB parameters, where $\kappa$'s are unknown
parameters.
Then we obtain
\be
\Delta \tilde{A}_{ij}^{q} (\Lambda_{C})&=&
\frac{A_{i}^{q}-A_{j}^{q}}{m_{\tilde{q}}} ~(\Lambda_{C}) =
\Delta \kappa_{ij}^{q}~
\frac{M_{2}(\Lambda)}{m_{\tilde{q}}(\Lambda_{C})}
=\Delta \kappa_{ij}^{q}~
\frac{M_{2}(\Lambda_{C})}{m_{\tilde{q}}(\Lambda_{C})}~
\frac{M_{2}(\Lambda)}{M_{2}(\Lambda_{C})},
\label{deltaAt2}
\ee
where
$\Delta \kappa_{ij}^{{q}}=\kappa_{i}^{q}-\kappa_{j}^{q}$
with $i,j=1,\dots,5$.
Similarly, we find
\be
\Delta \tilde{A}_{ij}^{\ell} (\Lambda_{C})
&=& \Delta \kappa_{ij}^{\ell}~
\frac{M_{2}(\Lambda_{C})}{m_{\tilde{\ell}}(\Lambda_{C})}
\frac{M_{2}(\Lambda)}{M_{2}(\Lambda_{C})}~~,i,j=2,4,5,
\label{deltaAt3}
\ee
and
\be
\Delta a^{q} &= & \Delta \kappa_{m_{q}}^{2}
\left[\frac{M_{2}(\Lambda_{C})}{m_{\tilde{q}}(\Lambda_{C})}\right]^{2}
\left[\frac{M_{2}(\Lambda)}{M_{2}(\Lambda_{C})}\right]^{2},~
\Delta a^{\ell}=  \Delta \kappa_{m_{\ell}}^{2}
\left[\frac{M_{2}(\Lambda_{C})}{m_{\tilde{\ell}}(\Lambda_{C})}\right]^{2}
\left[\frac{M_{2}(\Lambda)}{M_{2}(\Lambda_{C})}\right]^{2}.
\label{deltaa2}
\ee
In  (\ref{deltaa2}) we have suppressed the prefixes
$L$ and $R$, because they have the same form.
To be more explicit, we assume
\be
m_{E_{R}}(\Lambda_{C}) &\gsim& 100~\mbox{GeV}~\mbox{and}~
M_{2} (\Lambda_{C})\lsim  1~\mbox{TeV},
\ee
so that Eq.~(\ref{restriction})  implies that
$M_{2}(\Lambda)  \gsim  10~\mbox{GeV}$.
This means
\be
\frac{M_{2}(\Lambda)}{M_{2}(\Lambda_{C})}
& > &  10^{-2}.
\label{constraintM}
\ee
With this value of the enhancement factor,
we obtain
\be
\Delta \tilde{A}(\Lambda_{C})
 & \gsim &10^{-2}\Delta \kappa  ~\mbox{and}~
\Delta a(\Lambda_{C})   \gsim 10^{-4}\Delta \kappa^{2},
\label{inequal}
\ee
where we have assumed that
$m_{\tilde{\ell}} \simeq m_{\tilde{q}} \simeq 
M_{2}$ at $\Lambda_{C}$.
In Figs.~1 and 2 we show the IR attractiveness of $h_{D}/M_{2}Y_{D}$
and $m^{2}_{Q}/|M_{2}|^{2}$ for $\delta=1$.
The constraint  (\ref{constraintM}) with (\ref{enhancement}) implies that 
$\Lambda/\Lambda_{C} \lsim 5\times 10^{3}$ in this case.
Since between $\Lambda_{C}$ and $M_{\rm SUSY}$
there are only logarithmic corrections, we may assume that
the inequalities (\ref{inequal}) remain unchanged at
$M_{\rm SUSY}$. As we have seen in the previous section, 
$\Delta \tilde{A} \lsim 1$ and 
$\Delta a \lsim 1$ are sufficient (except for 
the fine tuning (\ref{A13}) to suppress 
the contributions to EDM of the neutron)
to suppress FCNCs and dangerous CP violating phases
at $M_{\rm SUSY}$. Therefore, at most an $O(100)$ disorder of the SSB
parameters at $\Lambda$  may be allowed.

Once again, $S_{3}$ symmetry can suppress
sufficiently FCNCs and CP phases that are generated by
 the SSB parameters in $O(1)$ disorder at $M_{\rm SUSY}$,
and the asymptotically free $SU(2)_{L}$ and $SU(3)_{C}$ gauge interactions
in extra dimensions can bring
 the SSB parameters in  $O(100)$ disorder 
 at the cutoff scale down
to those in $O(1)$ disorder at the compactification scale.
In this way, FCNCs and CP phases enjoy a double suppression.

\section{Conclusions}
A low energy nonabelian flavor symmetry is certainly a powerful tool
to suppress flavor-changing neutral currents (FCNCs) and CP violating phases
that are induced by soft supersymmetry breaking (SSB) terms.
In the  case of the MSSM, where 
only two types of Higgs supermultiplets $H^{U}$ and $H^{D}$ exist,
any nonabelian  flavor symmetry has to be explicitly broken to
describe experimental data.
Therefore, it would be unnatural to assume
a flavor symmetry only in the SSB sector.
However, if the Higgs sector is extended, and 
the Higgs supermultiplets belong to
a non-trivial representation of a flavor group,
phenomenologically viable possibilities arise.
We have considered the smallest nonabelian group
$S_{3}$ as a flavor group to extend the MSSM.
Under this flavor group, not only  the fermion multiplets, but
also the two types of the Higgs multiplets belong
to the three dimensional representation of $S_{3}$ \cite{kubo1,kubo2}.
The most general Higgs superpotential
has an accidental continuous global symmetry, implying
that a lot of pseudo Goldstone bosons will appear
after spontaneous symmetry breaking of the
electroweak symmetry \cite{kobayashi3}. 
To overcome this problem, we introduced
soft $S_{3}$ breaking B terms in the SSB sector,
because violation of $S_{3}$ can be confined in this sector.
We found that in this model
FCNCs and CP phases, which are induced by
the SSB parameters in $O(1)$ disorder at  $M_{\rm SUSY}$,
are  sufficiently suppressed to satisfy
the experimental constraints.

Then we extended  the $S_3$ invariant 
supersymmetric model   into $\delta$ extra dimensions 
compactified on an orbifold with a compactification scale 
$\Lambda_{C}=1/R$, and assumed that
the matter multiplets and the $U(1)_Y$ gauge multiplet are located 
at a fixed point, while the $SU(2)_L$ and $SU(3)_C$ gauge multiplets propagate
in the bulk. 
In this way, the $SU(2)_{L}$ and $SU(3)_{C}$ gauge couplings 
become asymptotically free, while the running of the $U(1)_{Y}$
coupling is frozen.
These asymptotically free gauge couplings 
amplify
the infrared attractiveness of the SSB
parameters when running  from 
the cutoff scale
down to the compactification scale  \cite{kubo6,choi1,choi2,kaji1}.
We found  (\ref{inequal}) that thanks to the
double suppression mechanism,
a disorder of two orders of magnitude
in the SSB parameters
at the cutoff scale
may be allowed
to satisfy experimental constraints on FCNC processes
and CP violating phenomena.

\acknowledgments
This work(Y.K. and J.K.) is supported by the Grants-in-Aid 
for Scientific Research 
from the Japan Society for the Promotion of Science
(No. 13135210). K.Y.C. is supported in part by the KOSEF ABRL Grant
to Particle Theory Research Group of SNU, the BK21 program of Ministry
of Education, and Korea Research Foundation Grant 
No. KRF-PBRG-2002-070-C00022. This work(H.M.L.) is supported by the
European Community's Human Potential Programme under contracts
HPRN-CT-2000-00131 Quantum Spacetime, HPRN-CT-2000-00148 Physics Across the
Present Energy Frontier and HPRN-CT-2000-00152 Supersymmetry and the Early
Universe. H.M.L. was supported by priority grant 1096 of the Deutsche
Forschungsgemeinschaft.

\appendix

\section{Renormalization group equations }
Here we give the $\beta$ functions.
The model in Sect. IV is a simple
embedding of  the supersymmetric extension 
of the $S_3$ invariant SM into extra dimensions compactified 
on an  orbifold \cite{antoniadis1,arkani1,dienes1},
where it is assumed that the matter 
multiplets  and the $U(1)_Y$ gauge multiplet are located 
at a fixed point, while the $SU(2)_L$ and $SU(3)_c$ gauge 
multiplets propagate
in the bulk. 
The dominance of the power running yields  the approximate 
$\beta$ functions that are given below\footnote{We note that,
even with the reduction of the number of  the KK modes
on the orbifold, the coupling
of the KK modes to the brane fields is larger than the one of the zero mode
by $\sqrt{2}$ due to the difference of normalization of mode functions.}.

\vskip 0.5cm
\underline{\bf Gauge sector:}
\vskip 0.5cm
\beqa{
16\pi^2\beta_{g_{1}}&=&
\frac{1}{16\pi^2}\left(\sum l(R)\right)g^3_1\label{power-gauge1}\\
16\pi^2\beta_{g_{2}}&=&
-\frac{1}{16\pi^2}C_2(SU(2))F_\delta^2g^3_2,\label{power-gauge2}\\
16\pi^2\beta_{g_{3}}&=&
-\frac{1}{16\pi^2}C_2(SU(3))F_\delta^2g_3^3,\label{power-gauge3}\\
16\pi^2\beta_{M_{1}}&=&
\frac{2}{16\pi^2}\left(\sum l(R)\right)g_1^2M_1\label{power-gaugino1}\\
16\pi^2\beta_{M_{2}}&=&
-\frac{2}{16\pi^2}C_2(SU(2))F_\delta^2g_2^2M_2,\label{power-gaugino2}\\
16\pi^2\beta_{M_{3}}&=&
-\frac{2}{16\pi^2}C_2(SU(3))F_\delta^2g_3^2M_3,\label{power-gaugino3}
}
where $F_\delta=X^{1/2}_\delta(R\Lambda)^{\delta/2}$ 
with $X_\delta=\pi^{\delta/2}/\Gamma(1+\delta/2)=2(\pi)$ for $\delta=1(2)$,
the quadratic Casimir is given by $C_2(SU(N))=N$,
and $\sum l(R)=39/5$.
As explained in section III,
we ignore logarithmic contributions except those from $g_1$.

\vskip 0.5cm
\underline{\bf Quark sector:}
\vskip 0.5cm
\be
16\pi^2\beta_{Y^D_k}&=&
Y^D_k\bigg(-\frac{7}{15}g_1^2-3g_2^2F_\delta^2-\frac{16}{3}g_3^2F_\delta^2\bigg),
\\
16\pi^2\beta_{Y^U_k}&=&
Y^U_k\bigg(-\frac{13}{15}g_1^2-3g_2^2F_\delta^2
-\frac{16}{3}g_3^2F_\delta^2\bigg),\\
16\pi^2\beta_{h^D_k}&=&
\frac{7}{15}g^2_1(2M_1Y^D_k-h^D_k)+3g^2_2F_\delta^2(2M_2Y^D_k-h^D_k) 
\nonumber \\
&+&\frac{16}{3}g^2_3F_\delta^2(2M_3Y^D_k-h^D_k),\\
16\pi^2\beta_{h^U_k}&=&
\frac{13}{15}g^2_1(2M_1Y^U_k-h^U_k)+3g^2_2F_\delta^2(2M_2Y^U_k-h^U_k) 
\nonumber \\
&+&\frac{16}{3}g^2_3F_\delta^2(2M_3Y^U_k-h^U_k),
\ee
where $k=1,2,3,4,5$.
\be 
16\pi^2(\beta_{m^2_Q})^i_j&=&\delta^i_j\bigg[-\frac{2}{15}|M_1|^2g^2_1-6|M_2|^2g^2_2F_\delta^2-\frac{32}{3}|M_3|^2g^2_3F_\delta^2\bigg],\\
16\pi^2(\beta_{m^2_{U_R}})^i_j&=&\delta^i_j\bigg[-\frac{32}{15}|M_1|^2g^2_1-\frac{32}{3}|M_3|^2g^2_3F_\delta^2\bigg], \\
16\pi^2(\beta_{m^2_{D_R}})^i_j&=&\delta^i_j\bigg[-\frac{8}{15}|M_1|^2g^2_1-\frac{32}{3}|M_3|^2g^2_3F_\delta^2\bigg]
\ee
where $i,j=1,2,3$.

\vskip 0.5cm
\underline{\bf Lepton sector:}
\vskip 0.5cm
\be
16\pi^2\beta_{Y^E_k}&=&
Y^E_k\bigg(-\frac{9}{5}g^2_1-3g^2_2F_\delta^2\bigg), \ \ 
k=2,4,5,\label{power-lepton-yukawa}\\
16\pi^2\beta_{Y^N_k}&=&Y^N_k\bigg(-\frac{3}{5}g^2_1-3g^2_2F_\delta^2\bigg),
\ \ k=2,3,4.
\ee
\be
16\pi^2\beta_{h^E_k}&=&
\frac{9}{5}g^2_1(2M_1Y^E_k-h^E_k)+3g^2_2F_\delta^2(2M_2Y^E_k-h^E_k),
\ \ k=2,4,5,\label{power-lepton-A-term}\\
16\pi^2\beta_{h^N_k}&=&
\frac{3}{5}g^2_1(2M_1Y^N_k-h^N_k)+3g^2_2F_\delta^2(2M_2Y^N_k-h^N_k),
k=2,3,4.
\ee
\be
16\pi^2(\beta_{m^2_L})^i_j&=&\delta^i_j\bigg[-\frac{6}{5}|M_1|^2g^2_1-6|M_2|^2g^2_2F_\delta^2\bigg],\label{power-slepton}\\
16\pi^2(\beta_{m^2_{E_R}})^i_j&=&
\delta^i_j\bigg[-\frac{24}{5}|M_1|^2g^2_1\bigg]
\label{betamER}
\ee
where $i,j=1,2,3$. 
\be
16\pi^2(\beta_{m^2_{N_R}})^I_J&=&\delta^I_J(4X_{N_2}+2X_{N_4}),\\
16\pi^2(\beta_{m^2_{N_R}})^3_3&=&2X_{N_3},\\
16\pi^2\beta_{M_1}&=&2M_1(4|Y^N_2|^2+2|Y^N_4|^2),\\
16\pi^2\beta_{M_3}&=&2M_3(2|Y^N_3|^2)
\ee
where $I,J=1,2$ and sneutrino masses have no power running.

\vskip 0.5cm
\underline{\bf Higgs sector:}
\vskip 0.5cm
\be
16\pi^2\beta_{\mu_k}&=&\mu_k\bigg[-\frac{3}{5}g^2_1-3g^2_2
F_\delta^2 \bigg], \ \ k=1,3, \\
16\pi^2\beta_{B_k}&=&\left\{\begin{array}{l}
\frac{3}{5}g^2_1(2M_1\mu_k-B_k)+3g^2_2F_\delta^2(2M_2\mu_k-B_k), \ \ k=1,3, \\
\frac{3}{5}g^2_1(-B_k)+3g^2_2F_\delta^2(-B_k), \ \ k=4,5,6,7. \end{array}
\right.
\ee
\be
16\pi^2(\beta_{m^2_{a}})^i_j=\delta^i_j\bigg[-\frac{6}{5}|M_1|^2g^2_1-6|M_2|^2g^2_2F_\delta^2\bigg]
\ee
where $a=H^U, H^D$ and $i,j=1,2,3$.


\end{document}